\documentclass[runningheads]{svmult}

\usepackage{makeidx}   
\usepackage{graphicx}  
\usepackage{subeqnar}  
\usepackage{multicol}  
\usepackage{physmubb}  
\makeindex             

\begin{document}

\newcommand{\lae}{\stackrel{<}{\sim}}
\newcommand{\gae}{\stackrel{>}{\sim}}

\def\Journal#1#2#3#4{{#1} {\bf #2}, #3 (#4)}

\def\EPJ{{\em E. Phys. J.} C}
\def\RMP{{\em Rev. Mod. Phys.}}
\def\IJMA{{\em Int. J. Mod. Phys.} A}
\def\PTP{{\em Prog. Theor. Phys.}}
\def\ARNPS{{\em Ann. Rev. Nucl. Part. Sci.}}
\def\PTP{{\em Prog. Theor. Phys.}}
\def\PRr{{\em Phys. Rev.}}
\def\MPLa{{\em Mod. Phys. Lett.}A}
\def\PRL{{\em Phys. Rev. Lett}}
\def\PLB{{\em Phys. Lett.}B}
\def\PRD{{\em Phys. Rev.}D}
\def\NPB{{\em Nucl. Phys.}B}
\def\ZPC{{\em Z. Phys.}C}

\title*{The Top Quark:\\ Experimental Roots and Branches of Theory\footnote{Talk presented at HCP2002, Karlsruhe, Germany, October 2002.}}
\toctitle{The Top Quark:
\protect\newline Experimental Roots and Branches of Theory}
%
%
\titlerunning{The Top Quark}
%
\author{Elizabeth H. Simmons}

\authorrunning{Elizabeth H. Simmons}
%
%

\maketitle              


\vspace{-.5cm}

\section{Introduction}

Even before the top quark was discovered by the CDF\cite{cdfdisc} and
D\O\cite{d0disc} experiments in 1995, several of its properties could
be deduced from those of its weak partner, the bottom quark.  The
increment in the measured\cite{qb} value of $R = \frac{\sigma(e^+e^-
\to hadrons)} { \sigma(e^+ e^- \to \mu^+\mu^-)} $ at the b threshold
agreed with the predicted $\delta R^{SM} = \frac{1}{3}$, confirming
$Q^b = - \frac{1}{3}$.  Likewise, data\cite{afb} on the front-back
asymmetry for electroweak b-quark production confirmed that $T_3^b =
-\frac{1}{2}$.  Therefore, the b quark's weak partner in the SM was
required to be a color-triplet, spin-$\frac{1}{2}$ fermion with
electric charge $Q = \frac{2}{3}$ and weak charge $T_3 = \frac{1}{2}$.

Such a particle is readily pair-produced by QCD processes involving
$q\bar{q}$ annihilation or gluon fusion.  At the Tevatron's $\sqrt{s}
= 1.8 - 2.0$ TeV, a 175 GeV top quark is produced 90\% through
$q\bar{q} \to t\bar{t}$ and 10\% through $gg \to t\bar{t}$; at the LHC
with $\sqrt{s} = 14$ TeV, the opposite will be true.  Eventually, a
Linear Collider (LC) with $\sqrt{s} \sim 350$ GeV can operate as a top
factory.

In Run I, each Tevatron experiment measured some top quark
properties in detail and took a first look at others.  As discussed in
the next section, our understanding of top as a distinct entiety is
rooted in these measurements.

\section{Experimental Roots}

\subsection{Mass}\label{subsec:mass}

The top quark mass has been measured\cite{tmasscdf,tmassd0} by
reconstructing the decay products of top pairs produced at the
Tevatron; the best precision comes from the lepton + jets decay
channel.  The combined measurement from CDF and D\O\ is $m_t = 174.3
\pm 5.1$ GeV. This implies the Yukawa coupling $\lambda_t =
2^{3/4} G_F^{1/2} m_t$ is $\sim$ 1, the only observed Yukawa coupling
of ``natural'' size.

\begin{figure}[tb]
\begin{center}
$\ $\hspace{-1cm}\scalebox{.33}{\includegraphics{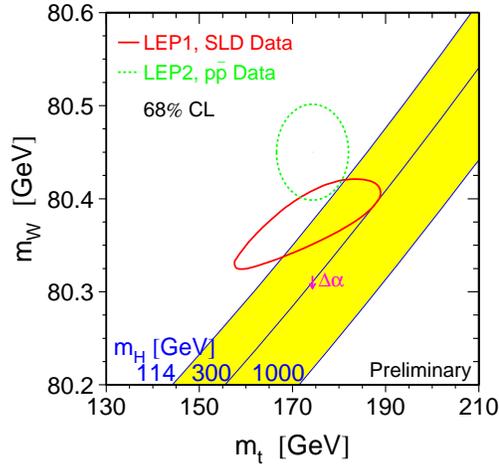}}
\end{center}
\vspace{-.7cm}
\caption[abh]{Predicted\cite{degrassi} $M_W(m_h, m_t)$ in the SM
compared\cite{lepewwg} to data on $M_W$, $m_t$.}
\label{fig:mwmtmh}
\end{figure}

The high precision with which the $m_t$ is known (comparable to that
of $m_b$ and better than that for light quarks \cite{pdg}) is useful
because radiative corrections to many precision electroweak
observables are sensitive to $m_t$.  Comparing the experimental
constraints on $M_W$ and $m_t$ with the SM prediction\cite{degrassi}
for $M_W (m_t, m_{Higgs})$ tests the consistency of the SM.  As Figure
\ref{fig:mwmtmh} shows, the data are suggestive, but not able to
provide a tightly-bounded value for $m_{Higgs}$.  Run II measurements
of the $W$ and top masses are expected\cite{wattshf8} to yield $\delta
M_W \approx$ 40 MeV (per experiment) and $\delta m_t \approx$ 3 GeV (1
GeV in Run IIb or LHC).  This should provide a much tighter
bound\cite{wattshf8} on the SM Higgs mass: $\delta M_H / M_H \leq
40$\%.

A measurement with $\delta m_t \approx 150$ MeV, can in principle
be extracted from near-threshold LC\cite{nlcmt} data on
$\sigma(e^+e^- \to t\bar{t})$.  The calculated line shape 
 rises at the remnant of what would have been the toponium 1S
resonance (see \ref{subsec:width}).  The location of the
rise depends on $m_t$; the shape and size, on the decay width
$\Gamma_t$. This technique has the potential for high precision
because it relies on counting color-singlet $t\bar{t}$ events,
making it insensitive to QCD uncertainties.

\begin{figure}[t]
\begin{center}
 \scalebox{.2}{\includegraphics{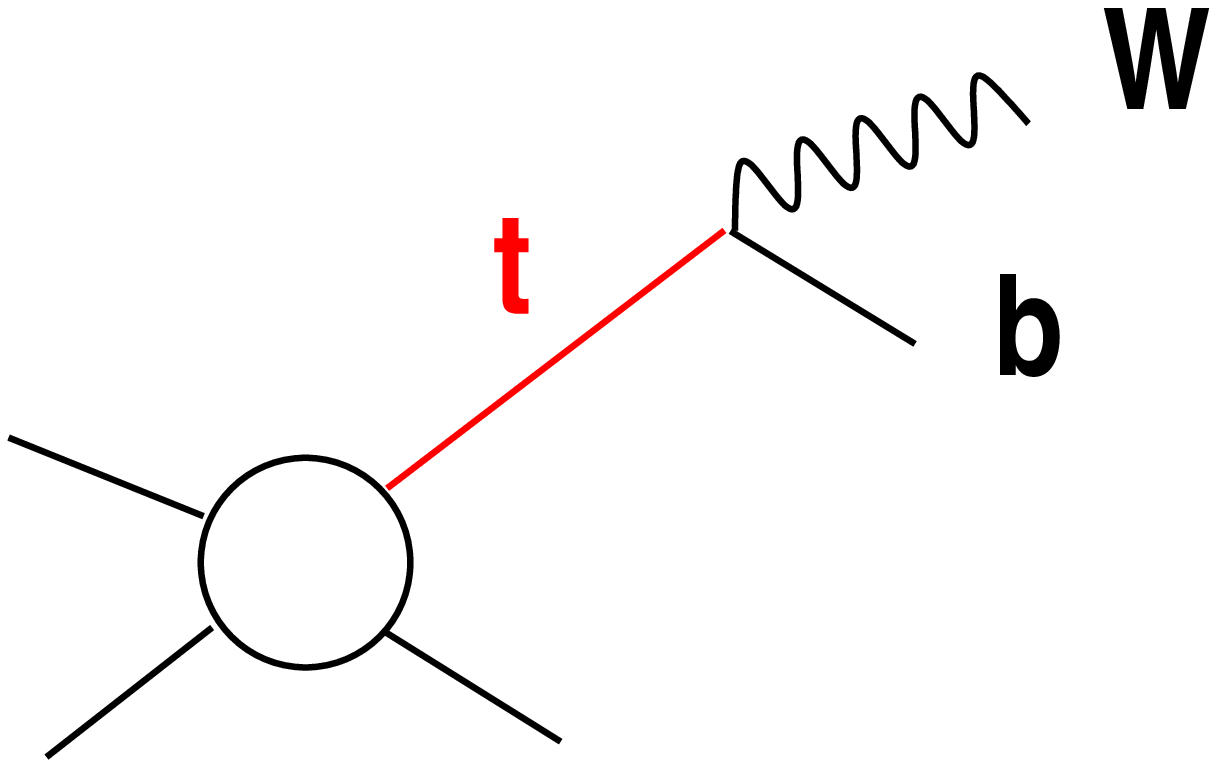}}\hspace{1cm}\scalebox{.2}{\includegraphics{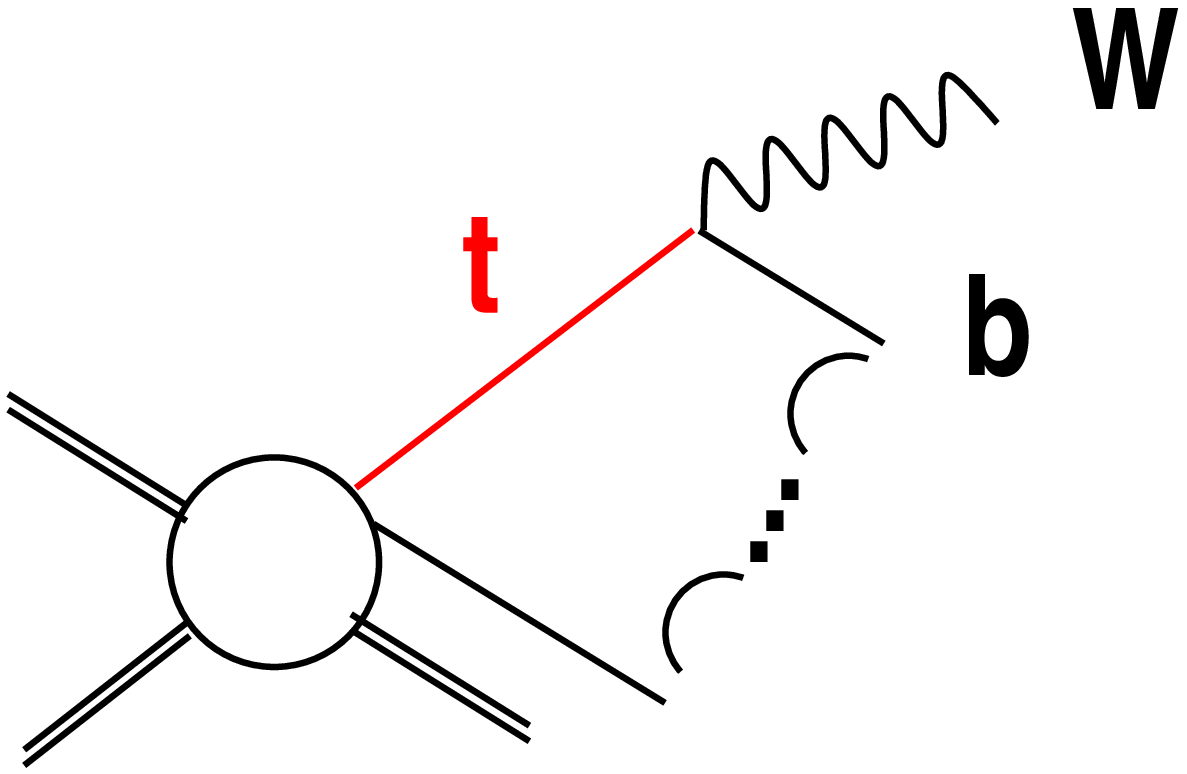}}
\end{center}
\caption[ab]{(left)Top production and decay. (right) Same, with b-quark
hadronization indicated.  After ref. \cite{willen}.}
\label{fig:massdiag}
\end{figure}

The definition of $m_t$ used to extract information from this data
must be chosen with care.  Consider, for example,
the mass in the propagator $D(p\!\!\!/) = i / (p\!\!\!/ - m_R -
\Sigma(p\!\!\!/))$.  In principle, one can reconstruct this mass from the
four-vectors of the top decay products, as is done in the Tevatron
measurements.  But this pole mass is inherently uncertain to ${\cal
O}(\Lambda_{QCD})$.  The top production and decay
process sketched in Figure \ref{fig:massdiag}(left) is, in reality,
complicated by QCD hadronization effects which connect the b-quark from top
decay to other quarks involved in the original scattering,\cite{willen} as
in Figure \ref{fig:massdiag}(right). 

\begin{figure}[tb]
\begin{center}
\vspace{.2cm}
{\hspace{0cm}\scalebox{.8}{\includegraphics*[40mm,205mm][100mm,250mm]{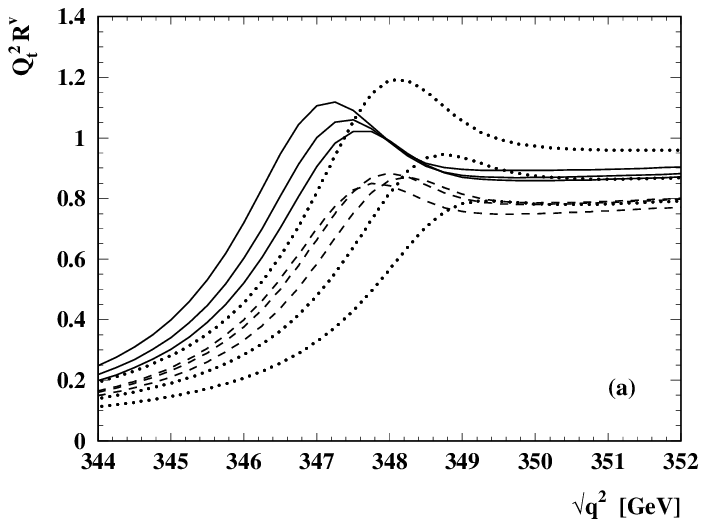}}} \hspace{1cm}
\hbox{\scalebox{.8}{\includegraphics*[40mm,205mm][100mm,250mm]{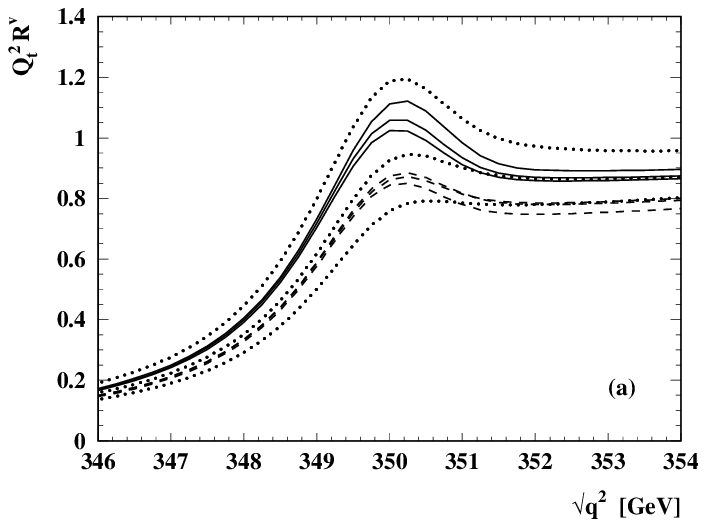}}}
\end{center}
\caption[ab]{Near-threshold cross-section for photon-induced top production
at an NLC\cite{hoang} calculated (left) in the pole mass scheme and (right)
in the 1S mass scheme.  Leading-order (dotted), NLO (dashed) and NNLO
(solid) curves are shown with renormalization scales $\mu$ = 15 (topmost),
30, and 60 GeV}
\label{fig:phoindtt}
\end{figure}

Using a short-distance mass definition avoids these
difficulties.  A convenient choice is the 1S
mass\cite{hoang} $m_{1S} = m_{pole} - \frac{2}{9}\alpha_s^2 m_{pole} +
... $ where $2 m_{1S}$ is naturally near the peak of $\sigma(e^+e^-
\to t\bar{t})$ .   Figure \ref{fig:phoindtt} compares
the photon-induced $t\bar{t}$ cross-section near threshold as
calculated in the pole mass and 1S mass schemes (for $m_t = 175$ GeV
and $\Gamma_t = 1.43$ GeV).  In the pole mass scheme, the location and
height of the peak vary with renormalization scale and order in
perturbation theory.  Using the short-distance mass
renders the peak location stable and large higher-order corrections
are avoided.

\subsection{Top Width and Decays}\label{subsec:width}

\smallskip
\noindent{\tt top decay width}
\smallskip

In the 3-generation SM, data on the lighter quarks combined with CKM
matrix unitarity implies\cite{pdg} ${ 0.9991 < \vert V_{tb} \vert <
0.9994}$.  Thus the top decays almost exclusively through $t \to W
b$. SM calculations including $m_b$, $m_W$ and radiative corrections
yield \cite{lhctop} $\Gamma_t/\vert V_{tb}\vert^2 = 1.42\, {\rm GeV}$
.  As a result, the top decays in $\tau_t \approx 0.4 \times 10^{-24}$
s, a time much less than $\tau_{QCD} \approx 3 \times 10^{-24}$ s. The
top quark therefore decays before it can hadronize.

 A precise measurement of the top quark width can be made at an LC
running at $\sqrt{s} \sim 350$ GeV by exploiting the fact that
$\Gamma_t$ controls the threshold peak height in $\sigma(e^+e^- \to
t\bar{t})$.  Recent results of NNLL calculations \cite{hoangnew}
suggest that the theoretical uncertainty will be only $\delta
\sigma_{tt}/\sigma_{tt} \approx 3\%$.

\smallskip
\noindent{\tt W helicity in top decay}
\smallskip

The SM predicts the fraction (${\cal F}_0$) of top quark decays to
longitudinal (zero-helicity) $W$ bosons will be large, due to the top
quark's big Yukawa coupling ${\cal F}_0 = \frac{m_t^2/2 M_W^2}{1 +
m_t^2/2 M_W^2} = (70.1 \pm 1.6)\%\ $.  One can measure ${\cal F}_0$ in
dilepton or lepton+jet events by exploiting the correlation of the $W$
helicity with the momentum of the decay leptons: a positive-helicity
$W$ (boosted along its spin) yields harder charged leptons than a
negative-helicity $W$.

\begin{table}[b]
\caption[]{Predicted\cite{tev2000} precision of Run II $W$ helicity
results for several $\int {\cal L} dt$.}
\begin{center}
\begin{tabular}{|l|c|c|c|}
\hline
& 1 fb$^{-1}$ & 10 fb$^{-1}$ & 100 fb$^{-1}$ \\
\hline
$\delta {\cal F}_0$ & 6.5\% & 2.1\% & 0.7\%\\
$\delta {\cal F}_+$ & 2.6\% & 0.8\% & 0.3\%\\
\hline
\end{tabular}
\end{center}
\label{tab:helic}
\end{table}

CDF has measured\cite{whelicitycdf} the lepton $p_T$ spectra for
dilepton and lepton + jet events.  By assuming no positive-helicity
$W$'s are present, CDF obtains the limit ${\cal F}_0 =
0.91\pm0.37\pm0.13$ (i.e., no more than 100\% of the W's are
longitudinal).  By setting ${\cal F}_0$ to its SM value of 0.70, they
obtain the 95\% c.l. upper limit ${\cal F}_+ < 0.28$ (i.e., no more
than $1 - {\cal F}_0$ have positive helicity).  More informative
constraints are expected from Run II (see Table \ref{tab:helic}).

\smallskip
\noindent{\tt $b$ quark decay fraction}
\smallskip

The top quark's decay fraction to $b$ quarks be measured by
CDF\cite{bbmeascdf} to be $B_b \equiv {\Gamma(t \to b W)} / {\Gamma(t \to q
W)} = 0.99 \pm 0.29$.  In the three-generation SM, 
\begin{equation}
B_b \equiv {\vert V_{tb}\vert^2 \over {\vert V_{tb}\vert^2 + \vert V_{ts}\vert^2
 + \vert V_{td}\vert^2}} \, .
\label{bbref}
\end{equation}
Three-generation unitarity dictates that the denominator of (\ref{bbref}) is
1.0, so that the measurement of $B_b$ implies\cite{bbmeascdf}
$\vert V_{tb} \vert > 0.76$ at 95\% c.l.
However, within the 3-generation SM, data on the light quarks
combined with CKM unitarity has already provided\cite{pdg} 
the much tighter constraints $ 0.9991 < \vert V_{tb} \vert < 0.9994$.

If we add a fourth generation of quarks, the analysis differs.  
D\O\ has constrained\cite{pdg} any 4-th generation $b^\prime$ quark to have
a mass greater than $m_t - m_W$; i.e., top cannot 
decay to $b^\prime$.  Then (\ref{bbref}) remains valid, but the denominator of the RHS no longer need equal 1.0.  The CDF
measurement of $B_b$ now implies $ \vert V_{tb} \vert \gg \vert V_{td} \vert\,,
\, \vert V_{ts} \vert$, a stronger constraint than that available \cite{pdg} from 4-generation CKM unitarity:  $0.05 < \vert V_{tb} \vert < 0.9994$. 

Direct measurement of $\vert V_{tb} \vert$ in single top-quark production
(via $q\bar{q} \to W* \to t\bar{b}$ and $gW \to t\bar{b}$ as in Figure \ref{fig:singtdiag}) at the Tevatron
should reach an accuracy\cite{tev2000} of 10\% in Run IIa (5\% in Run IIb).

\smallskip
\noindent{\tt FCNC decays}
\smallskip

\begin{figure}[bt]
$\ $\hspace{-1cm}
\scalebox{.35}{\includegraphics{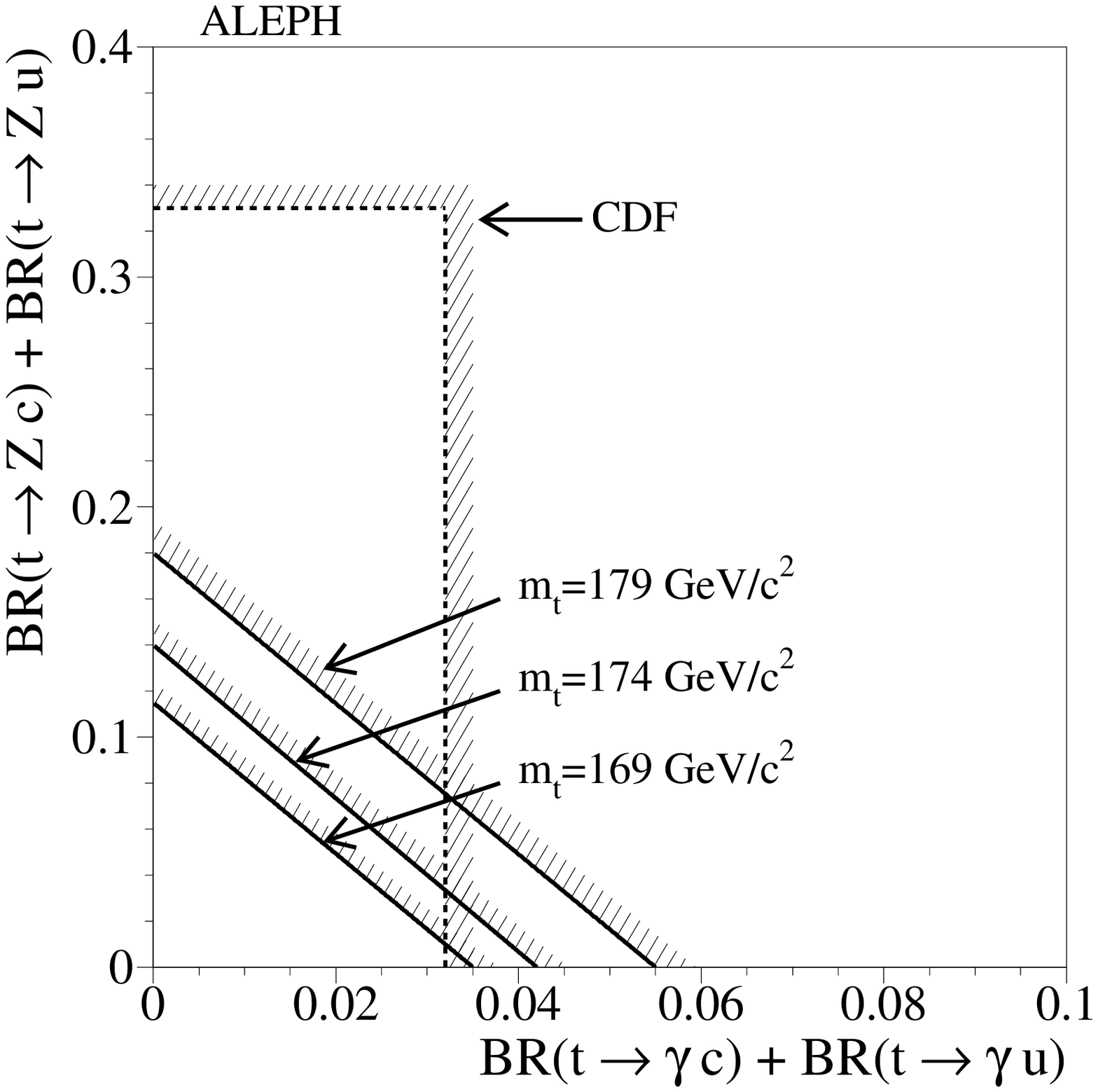}}  
\hspace{0cm} 
\raisebox{-35pt}{\scalebox{.32}{\includegraphics{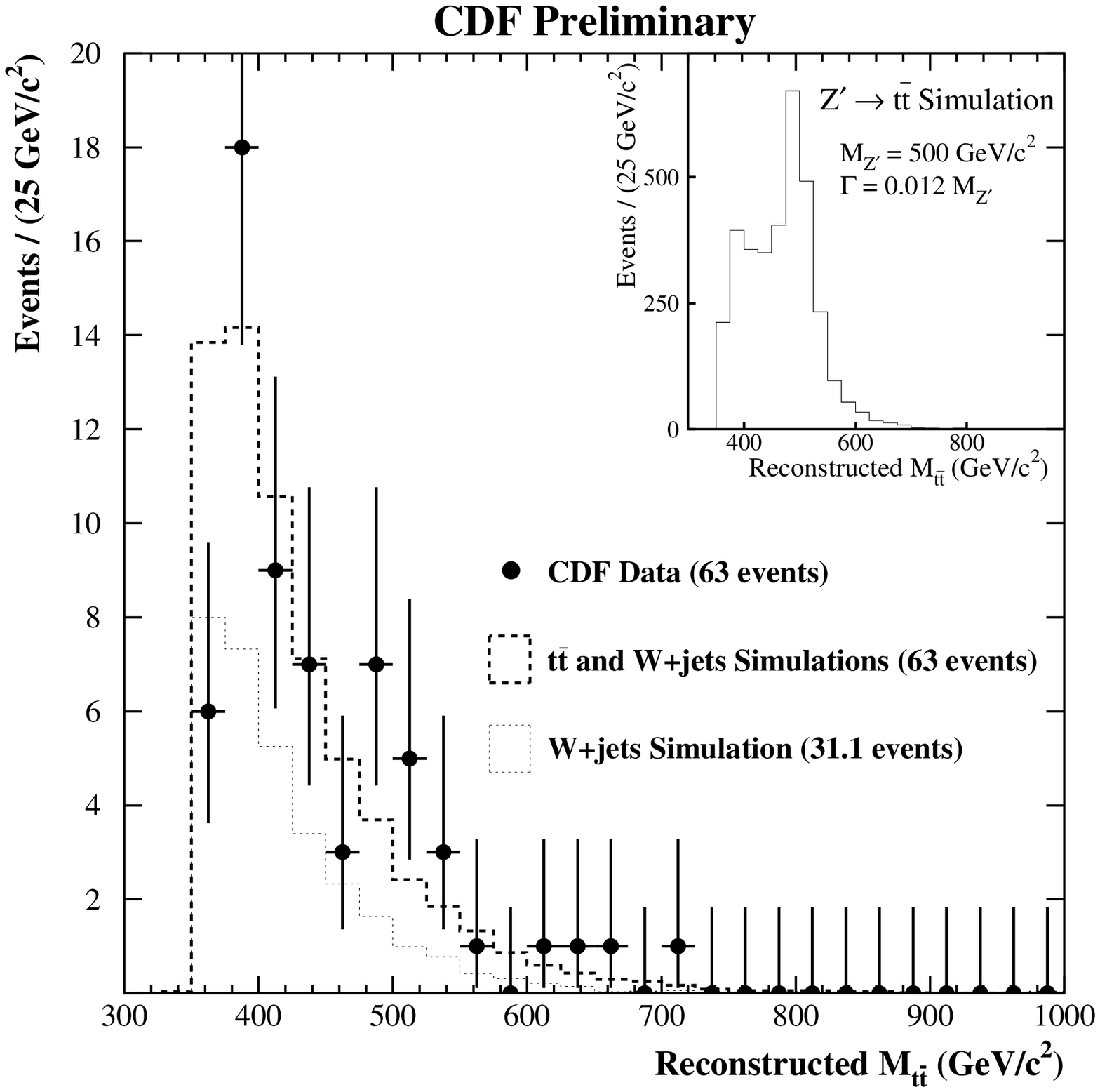}}}
\vspace{-1.5cm}
\caption{(left) FCNC limits from CDF \cite{fcnccdf}and ALEPH\cite{fcncaleph}.  (right) CDF
\cite{leptophobe} limits on leptophobic Z'}
\label{fig:fcnczp}
\end{figure}

CDF\cite{fcnccdf} and ALEPH\cite{fcncaleph} have set complementary limits on the flavor-changing neutral decays $t\to \gamma q$ and $t\to Zq$ as shown in \ref{fig:fcnczp}.  Run II will provide increased sensitivity to these channels\cite{tev2000} as indicated in Table
\ref{tab:fcnc}.

\begin{table}[tb]
\caption{Run II sensitivity\protect\cite{tev2000} to FCNC top decays as a
function of $\int {\cal L}dt$. }
\begin{center}
\begin{tabular}{|l|c|c|c|}
\hline
 & 1 fb$^{-1}$ & 10 fb$^{-1}$ & 100 fb$^{-1}$ \\
\hline 
$BR(t\to Z q)$  & 0.015 & $3.8\times 10^{-3}$ & $6.3\times 10^{-4}$ \\
$BR(t\to \gamma q)$ & $3.0\times 10^{-3}$ & $4.0\times 10^{-4}$ & $8.4\times 10^{-5}$\\
\hline
\end{tabular}
\end{center}
\label{tab:fcnc}
\end{table}

\subsection{Pair Production}\label{subsec:pairprod}

The top pair production cross-section has been measured by
CDF\cite{tcrosscdf} and D\O \cite{tcrossd0}, again, most precisely in
the lepton + jets channel.  The combined average of
$\sigma_{tt}(m_t=172\, {\rm GeV}) = 5.9 \pm 1.7$ pb is consistent with
SM predictions including radiative corrections.\cite{tcrosstheory}

Initial measurements of the invariant mass ($M_{tt}$) and transverse
momentum ($p_T$) distributions of the produced top quarks have yielded
preliminary limits on new physics. As seen in figure
\ref{fig:fcnczp}, e.g., a narrow 500 GeV Z' boson is inconsistent with
the observed shape of CDF's $M_{tt}$ distribution.\cite{leptophobe}

\begin{figure}[tb]
\scalebox{.35}{\includegraphics*[0mm,50mm][200mm,230mm]{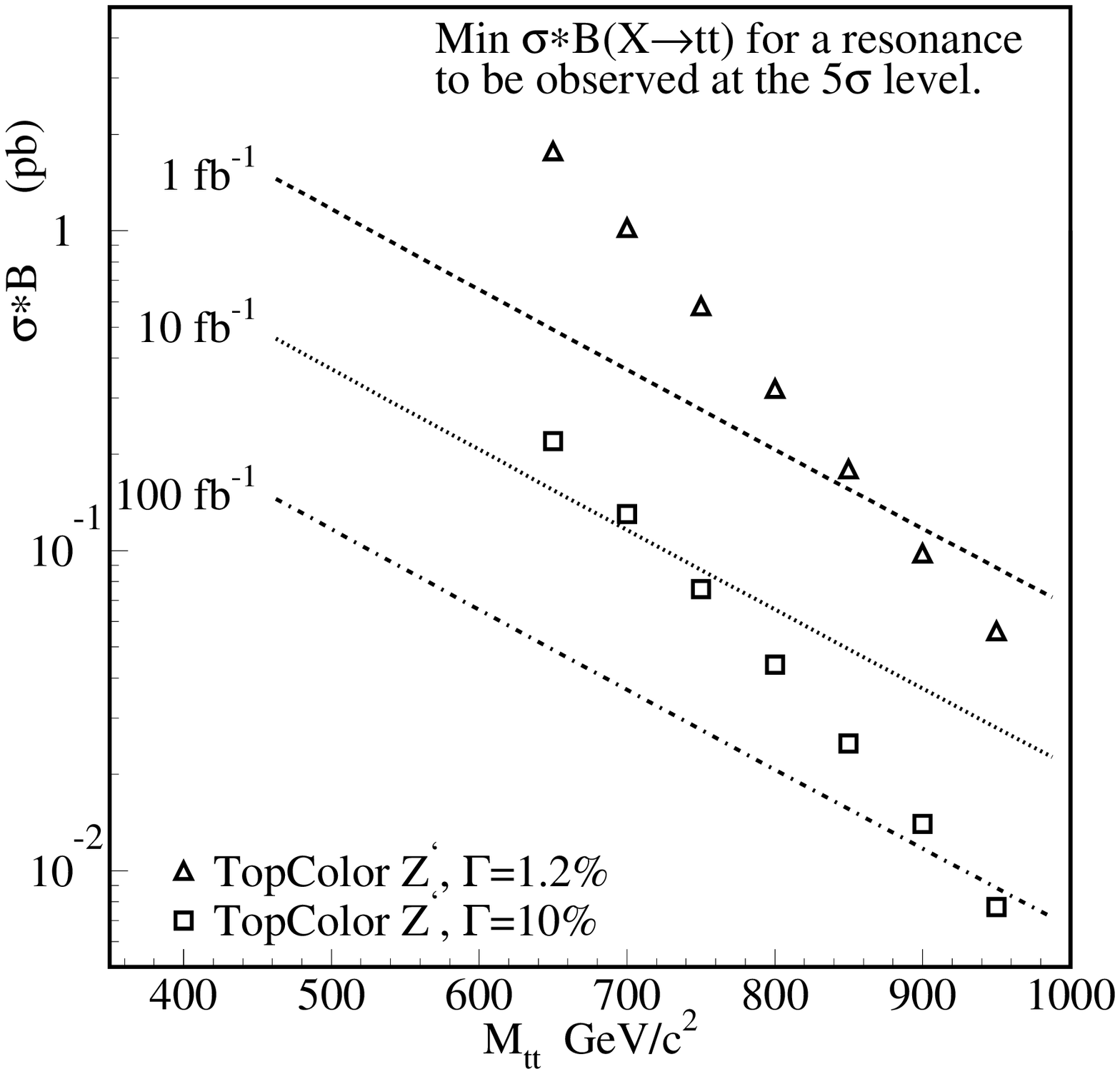}}
\hspace{1cm}
\raisebox{100pt}{\vbox{{\scalebox{.2}{\includegraphics{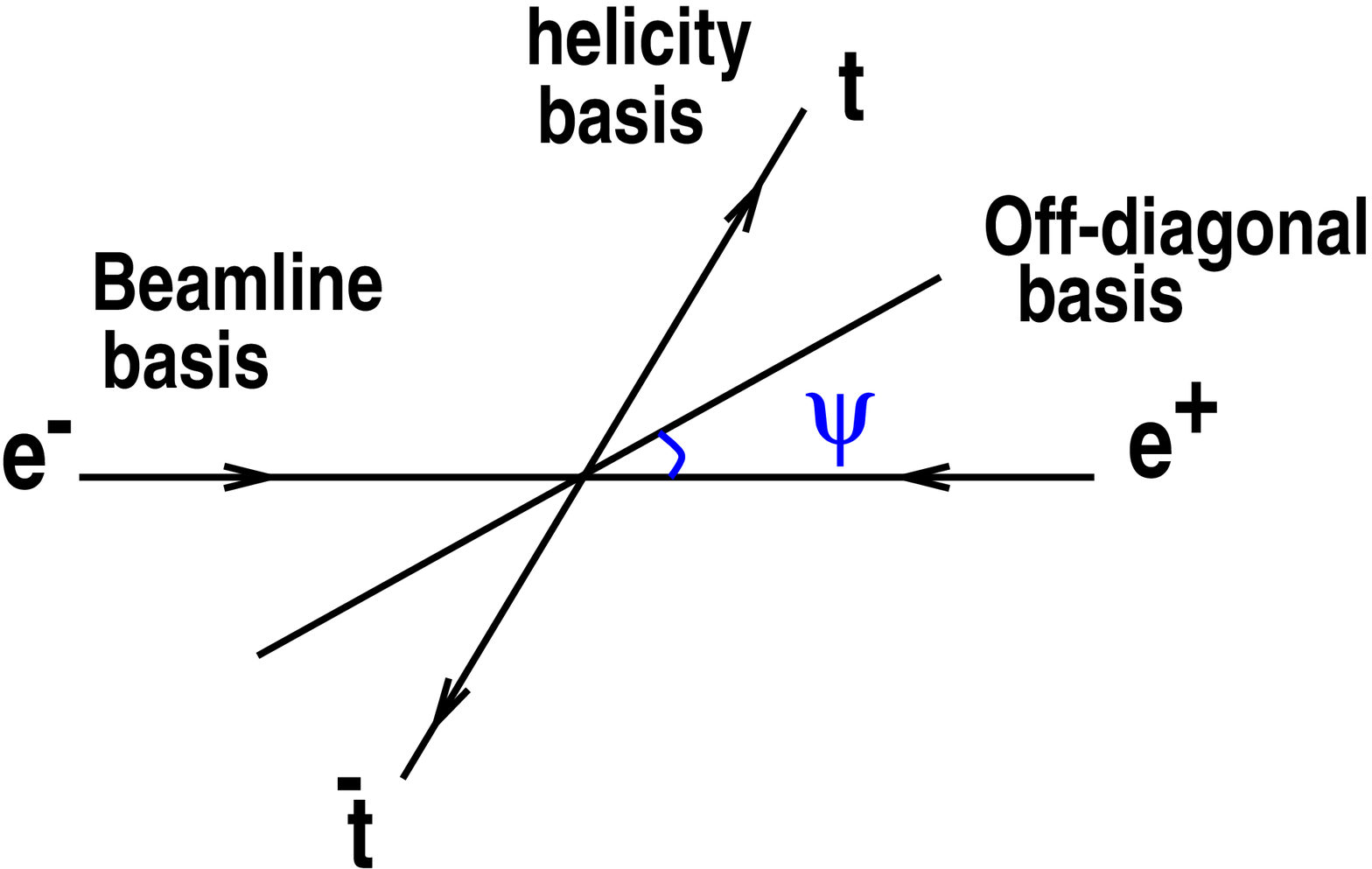}}}
\hspace{-4cm}\raisebox{-70pt}{\scalebox{.22}{\includegraphics{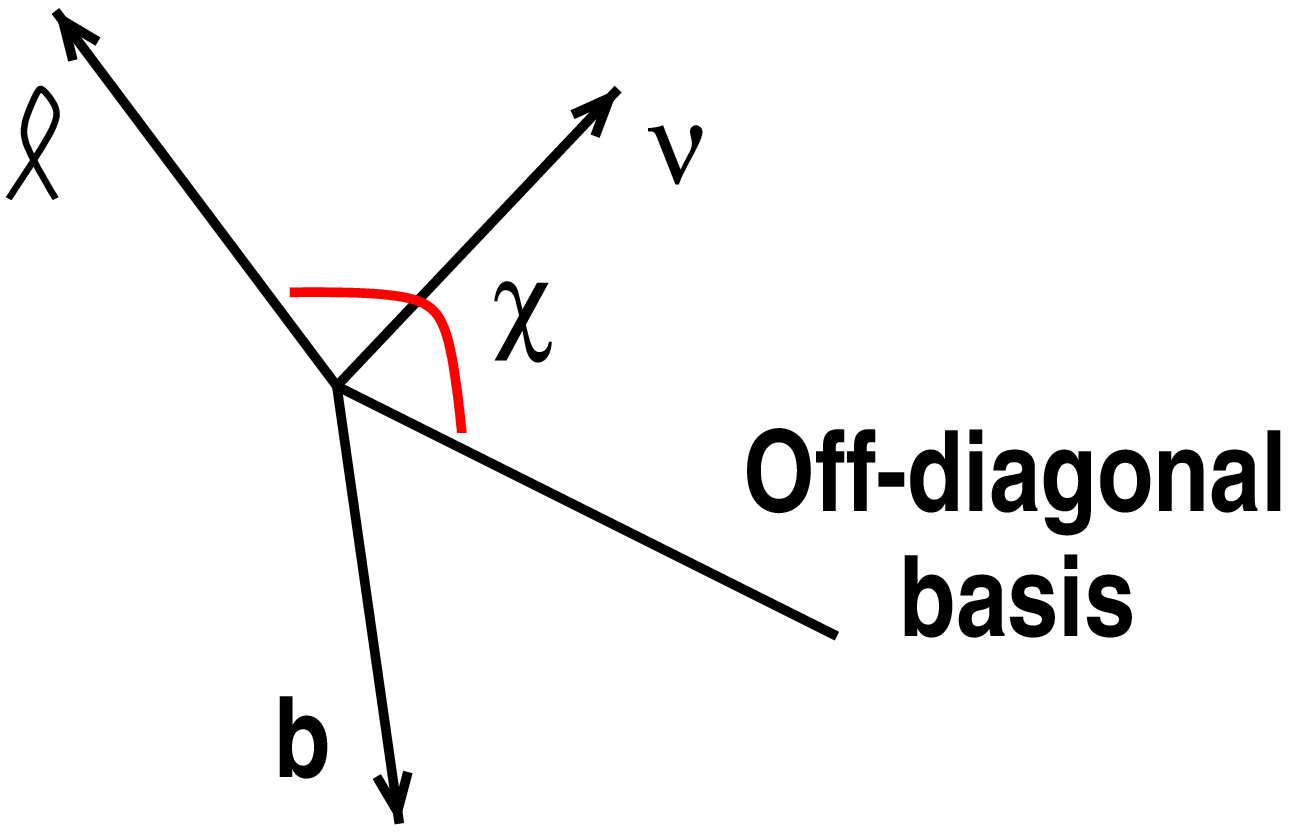}}}}}
\vspace{-.2cm}
\caption[ab]{(left)Anticipated\protect\cite{tev2000} Run II limits on
$\sigma\cdot B(X \to t\bar{t})$ 
(right) Definitions of the off-diagonal basis and decay lepton angles
for studying top spin correlations.}
\label{fig:sigbxtt}
\end{figure}

In Run II, the $\sigma_{tt}$ measurement will be dominated by
systematic uncertainties; the large data sample will be used to reduce
reliance on simulations.\cite{tev2000} An integrated luminosity of 1
(10, 100) fb$^{-1}$ should \cite{tev2000} enable $\sigma_{tt}$ to be
measured to $\pm$ 11 (6, 5) \%.  The $M_{tt}$ distribution will then
constrain $\sigma\cdot B$ for new resonances decaying to $t\bar{t}$ as
illustrated in Figure \ref{fig:sigbxtt}.

\subsection{Spin Correlations}\label{subsec:spincorr}

When a $t\bar{t}$ pair is produced, the spins of the two fermions are
correlated.\cite{bargero}  Because the top quark decays before its spin can
flip,\cite{bigi}  the spin correlations between $t$ and $\bar{t}$
yield angular correlations among their decay products.  If $\chi$
is angle between the top spin and the momentum of a given decay product,
the differential top decay rate (in the top rest frame) is
\begin{equation}
\frac{1}{\Gamma} \frac{d \Gamma}{d \cos\chi} = 
\frac{1}{2} (1 + \alpha \cos\chi)
\end{equation}
The factor $\alpha$ is \cite{alphaval}  1.0 (0.41, -0.31, -0.41)
if the decay product is $\ell$ or $d$ ($W$, $\nu$ or $u$, $b$). 
Thus, dilepton events are best for
studying $t\bar{t}$ spin correlations.

At the Tevatron, spin correlations are best studied in an optimal
 ``off-diagonal'' basis\cite{parkeshadmi,mahlonparke} in which the
 spins are purely anti-correlated at leading order ($t_\uparrow
 \bar{t}_\downarrow + t_\downarrow \bar{t}_\uparrow$). The projection
 axis is identified by angle $\psi$
\begin{equation}
\tan\psi = \frac{\beta^2 \sin\theta^* \cos\theta^*}{1 - \beta^2
\sin^2\theta^*}
\end{equation}
where $\beta$ ($\theta*$)is the top quark's speed (scattering angle)
in the center-of-momentum scattering frame (see Figure
\ref{fig:sigbxtt}).  Writing the
differential cross-section in terms of the angles $\chi^\pm$ of the
decay leptons $\ell^\pm$,
\begin{equation}
\frac{1}{\sigma} \frac{d^2 \sigma}{d(\cos\chi_+) d(\cos\chi_-)} =
\frac{1}{4} (1 + \kappa \cos\chi_+ \cos\chi_-)
\label{eq:difsig}
\end{equation}
one finds $\kappa \approx 0.9$ in the SM for $\sqrt{s} = 1.8$ TeV.
D\O\ used the six Run I dilepton events to set\cite{spincorrd0} the
68\% c.l. limit $\kappa \geq -0.25$.  Run IIa promises $\sim$150 dilepton
events.\cite{tev2000}

\begin{figure}[tb]
\centerline{\scalebox{.2}{\includegraphics{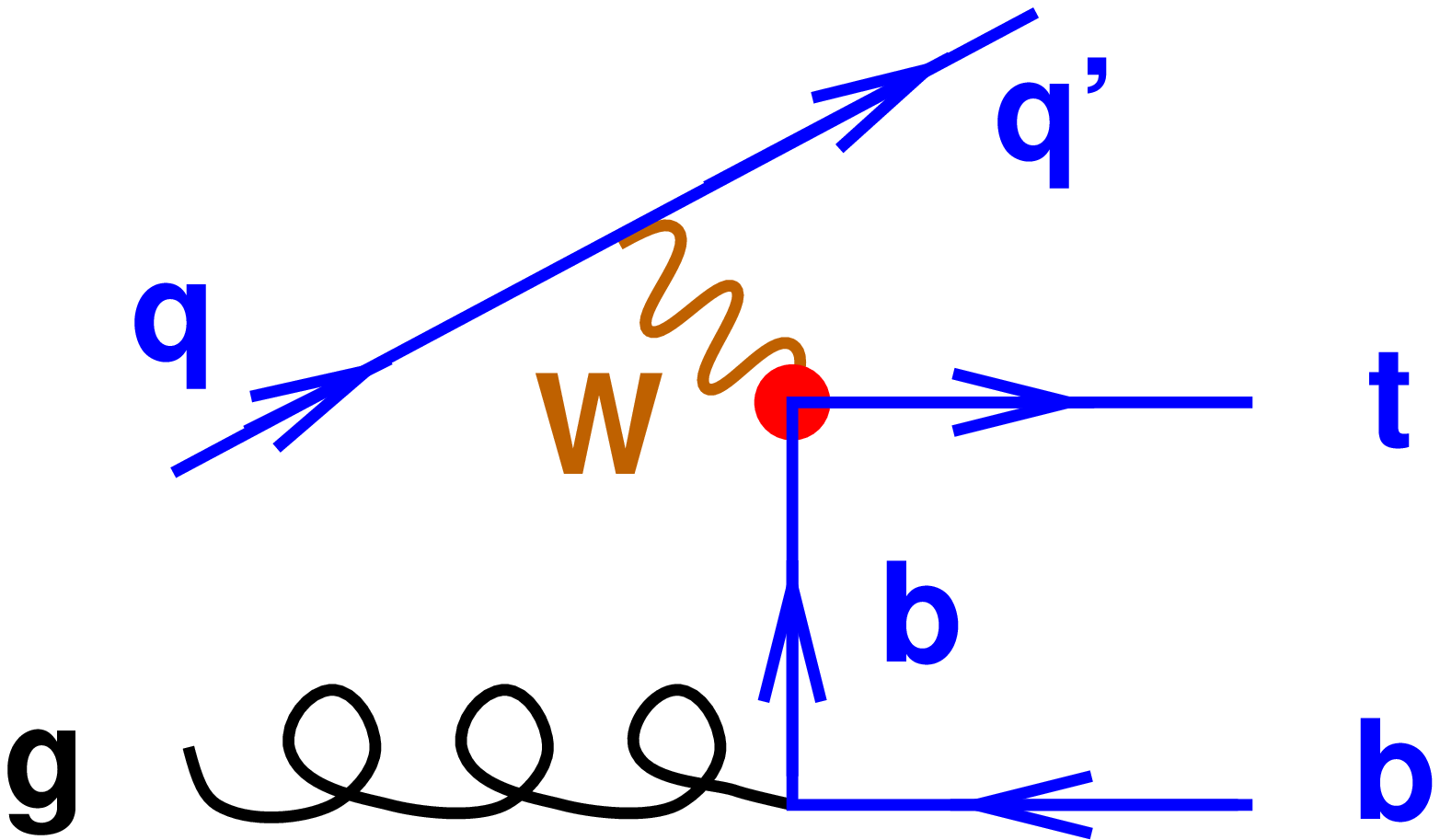}}
\hspace{.8cm}\scalebox{.22}{\includegraphics{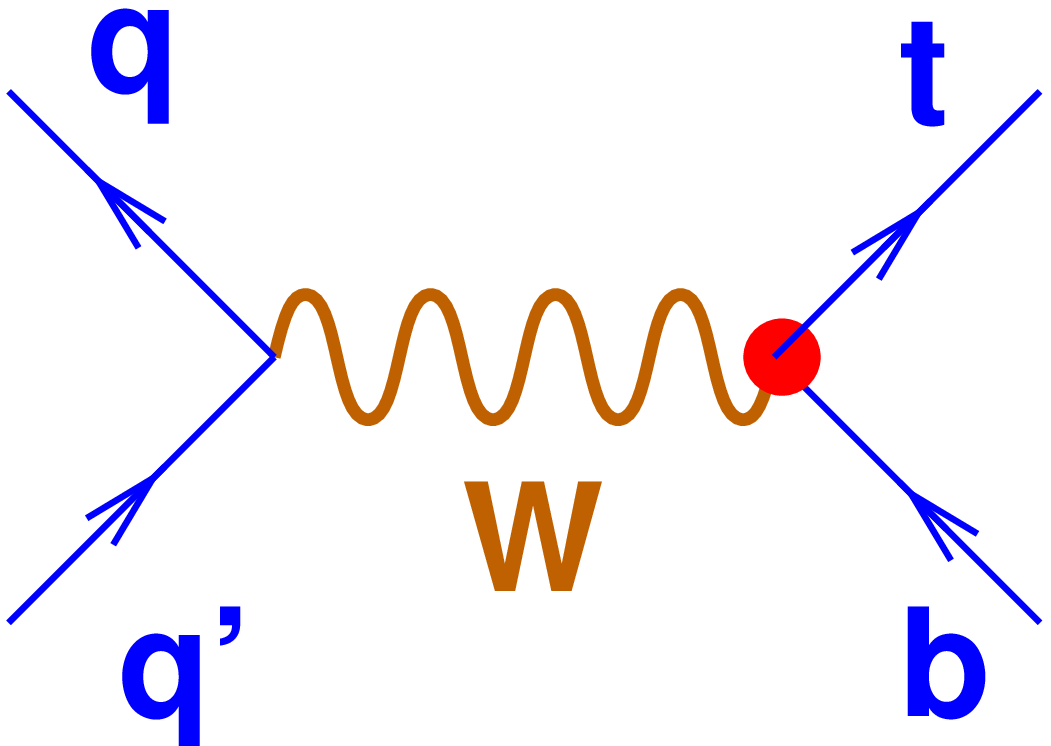}}
\hspace{.8cm}\scalebox{.2}{\includegraphics{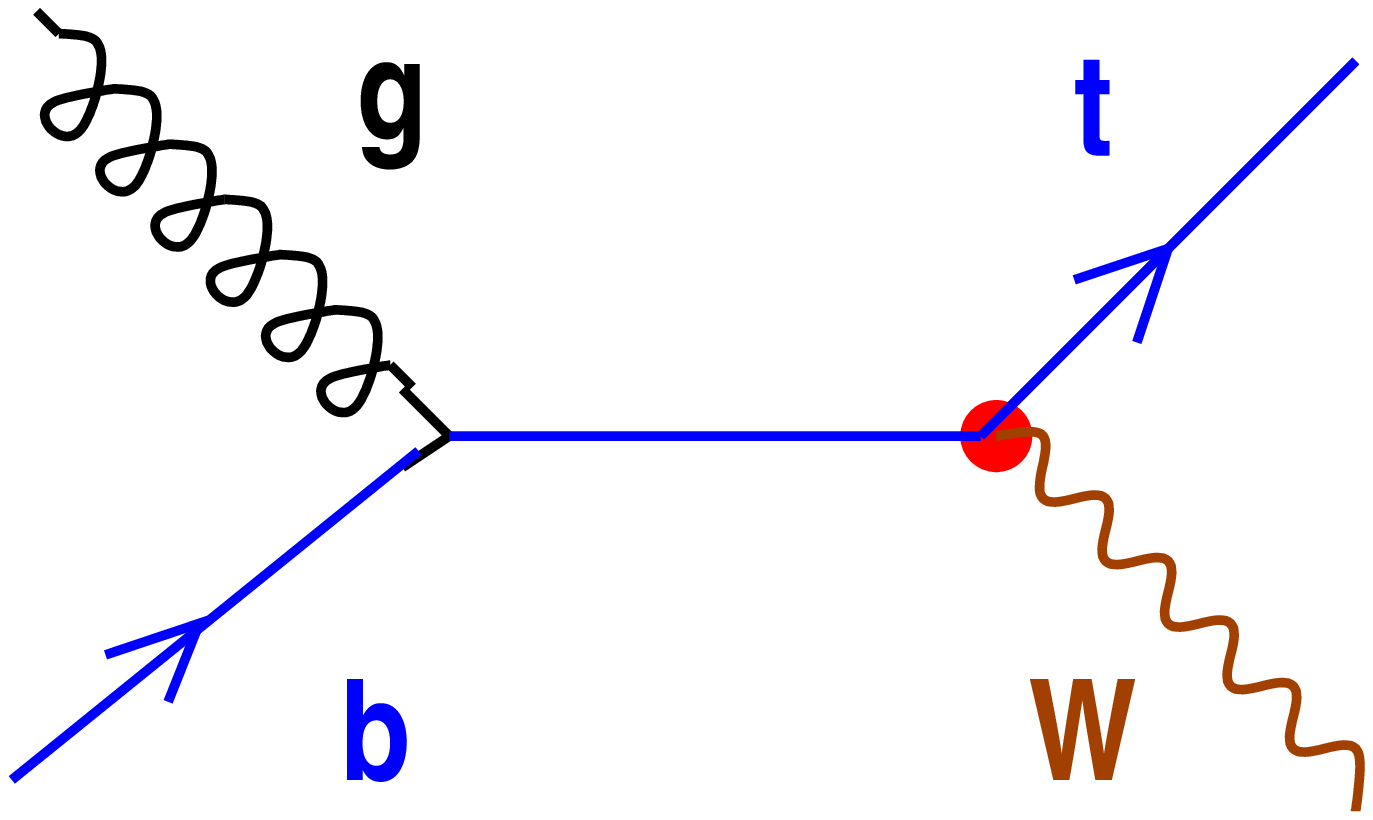}}}
\caption[aan]{Feynman diagrams for single top quark production.}
\label{fig:singtdiag}
\end{figure}

At the LHC, the top dilepton sample will be of order $4\times 10^5$
events per year\cite{lhctop} -- but no spin basis with nearly 100\%
correlation at all $\beta$ has been identified.  Near threshold,
angular momentum conservation favors like helicities; far above
threshold, helicity conservation favors opposite
helicities.\cite{lhctop} Spin effects in single top production at LHC
are also under study.  Promising variables include the angles between
the initial $d$ and decay $e$ (in the $W*$ process or $Wg$ fusion -
see Figure \ref{fig:singtdiag}) or the angles between the $e^+$ from
top decay and the $e-$ from W decay (in $gb \to tW$).

\section{Branches of Theory}

  We now explore how top influences some branches of theory beyond the
  SM. The small Run I top sample leaves open many intriguing
  possibilities.

\subsection{Light Neutral Higgs in MSSM}\label{subsec:lthiggs}

Radiative corrections to $m_h$ involving virtual top quarks introduce
a dependence on $m_t$; the heavy top makes these
significant\cite{susyrev} .  For $\tan\beta > 1$,
\begin{equation}
M_h^2 < M_Z^2 \cos^2(2\beta) + {3 G_f
    \over{\sqrt{2}\pi^2}}\ {m_t^4}\ ln
\left({\tilde m^2 \over m_t^2}\right)
\end{equation}
Including higher-order corrections, the most general limit appears to
be $M_h < 130$ GeV, well above the current bounds but in reach of
upcoming experiment.  This is comparable to the generic SUSY limit
$M_h < 150$ GeV from requiring the Higgs self-coupling to be
perturbative up to $M_{Planck}$. \cite{susyrev}

\subsection{Charged Higgs}\label{subsec:twohiggs}

\begin{figure}[tb]
$\ $ \hspace{1.5cm}
\raisebox{95pt}{\scalebox{.475}{\includegraphics*[100mm,100mm][140mm,166mm]{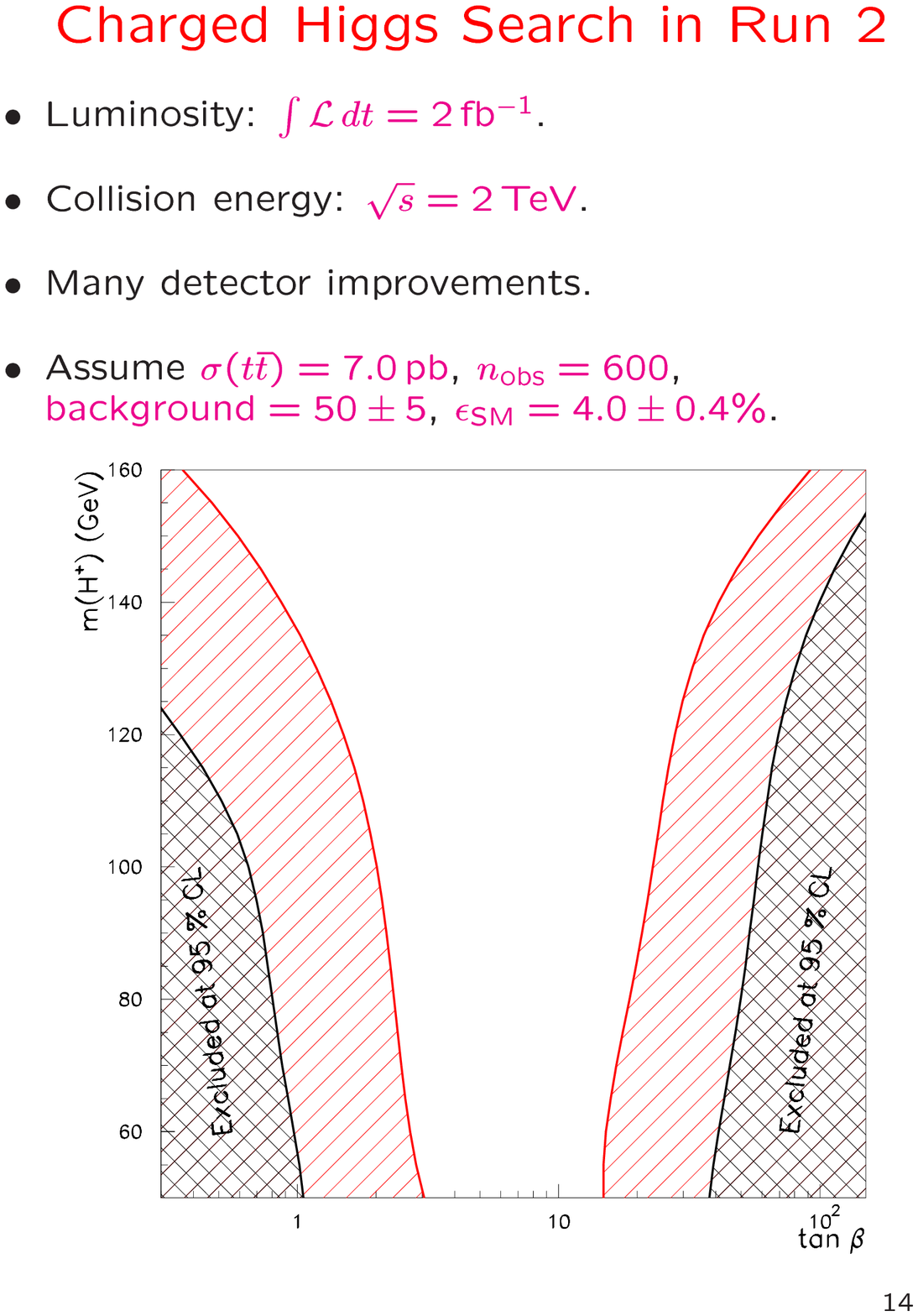}}}
\hspace{1.5cm}
\scalebox{.35}{\includegraphics{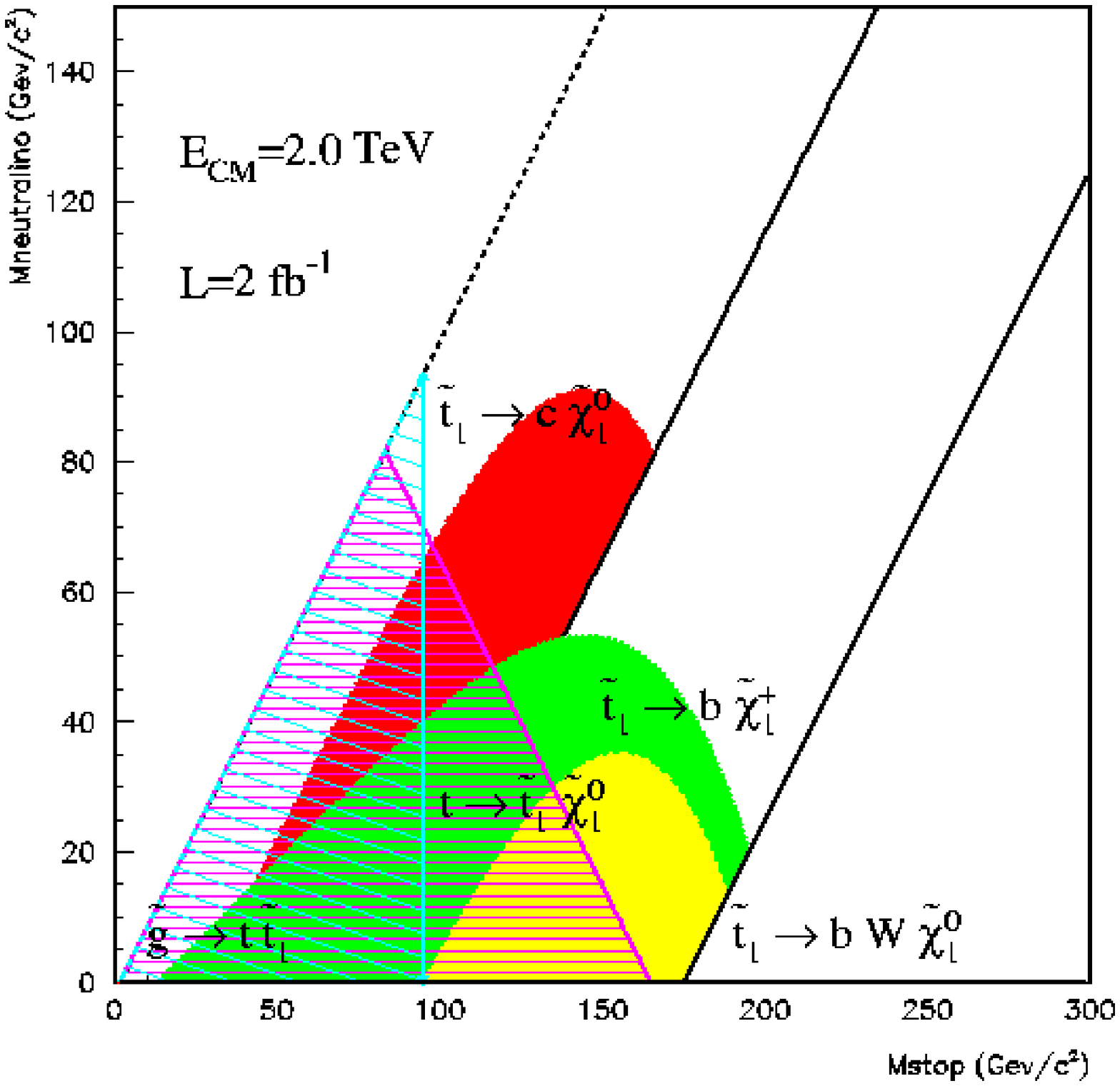}}
\vspace{-.25cm}
\caption{Projected D\O\ reach in Run II searches for (left) charged scalars in  $t\to
H^\pm b$ \protect\cite{tampere} assuming
$\sqrt{s} = 2$ TeV, $\int{\cal L} dt = 2 {\rm fb}^{-1}$, and
$\sigma(t\bar{t})$ = 7pb (right) stop searches in 
various channels.\protect\cite{run2stop}}
\label{fig:chgddh}
\end{figure}

Models from SUSY to dynamical symmetry breaking \cite{dsbrev} include
 light charged scalar bosons, into which top may decay: $t \to H^+ b$.
 If $m_H^\pm < m_t - m_b$, the decay $t \to H^+ b$ competes with the
 standard decay mode $t \to W b$.  Since the $t b H^\pm$ coupling
 depends on $\tan\beta$ as\cite{higgshunt}$
[m_t\cot\beta(1+\gamma_5) + m_b\tan\beta(1-\gamma_5)]$,   
the additional decay mode is significant for either large or small
values of $\tan\beta$.  The charged scalar decays as $H^\pm
\to c s$ or $H^\pm \to t^*b \to Wbb$ if $\tan\beta$ is small and as
$H^\pm \to \tau \nu_\tau$ if $\tan\beta$ is large.  
The final state reached through an intermediate $H^\pm$ will cause the
original $t\bar{t}$ event to fail the usual cuts for the lepton + jets
channel.  A reduced rate in this channel signals the
presence of a charged scalar.  Expected D\O\ limits in run II are shown in Figure \ref{fig:chgddh}.

\subsection{Sfermion Masses}\label{subsec:susy}

SUSY models must explain why the scalar Higgs boson acquires a
negative mass-squared (breaking the electroweak symmetry) while the
scalar fermions do not (preserving color and electromagnetism).  In
the MSSM with GUT unification or models with dynamical SUSY breaking,
the answer involves the heavy top quark.\cite{susyrev} In these
theories, the masses of the Higgs bosons and sfermions are related at
a high energy scale $M_\chi$:
\begin{equation}
M_{h,H}^2 (M_\chi) = m_0^2 + \mu^2\ \ \ \ \ \ \ \ \ \ \ \ M_{\tilde{f}}^2
(M_\chi) = m_0^2
\end{equation}
where the squared masses are all positive so that the vacuum preserves the
color and electroweak symmetries.  Renormalization group
running yields scalar masses at lower energies.\cite{ref39} 
  At scale $q$, the solution for $M_h$ is, e.g.,
\begin{equation}
M_h^2 (q) \approx M_h^2 (M_X) - {3\over{8\pi^2}} 
{\lambda_t^2} { \left( {\tilde{M}_{Q^3_L}}^2 +
{\tilde{M}_{t_R}}^2 + M_h^2 + A_{o,t}^2\right)} ln \left({M_X \over q }
\right) 
\end{equation}
and the top Yukawa coupling 
$\lambda_t$ is seen to be reducing $M_h^2$.  For $m_t \sim 175$ GeV,
this effect drives only the Higgs mass negative, 
just as desired.\cite{ref41}

SUSY models include bosonic partners for $t_L$ and $t_R$.  The
mass-squared matrix for the top squarks \cite{susyrev} (in the
$\tilde{t}_L, \tilde{t}_R$ basis)
\begin{equation}\null\hspace{-1cm}\tilde{m}_t ^2 = 
\pmatrix{\tilde{M}^2_Q + m_t^2 &\ & 
{m_t}(A_t + \mu\cot\beta)\cr 
+ M_Z^2(\frac12 - \frac23 \sin^2\theta_W)\cos2\beta &\ &\cr
\ &\ &\ \cr  {m_t}(A_t +
    \mu\cot\beta)& &\tilde{M}^2_U + m_t^2 \cr
&\ & + \frac23 M_Z^2 \sin^2\theta_W
    \cos2\beta\cr} 
\end{equation}
has off-diagonal entries proportional to $m_t$.  Hence, a
large $m_t$ can drive one of the top squark mass eigenstates to be
relatively light.  Experiment still allows a light stop,\cite{run1stop}; Run II will be sensitive to higher
stop masses in several decay channels\cite{run2stop} (Figure
\ref{fig:chgddh}).

Perhaps some of the Run I ``top'' sample included top
squarks.\cite{lightstop} If $m_{\tilde{t}} > m_t$, perhaps
$\tilde{t}\tilde{t}$ production occurred in Run I, with the top
squarks decaying to top plus $\tilde{N}$ or $\tilde{g}$
(depending on the masses of the gauginos).  If $m_t > m_{\tilde{t}}$
some top quarks produced in $t\bar{t}$ pairs in Run I may have decayed
to top squarks via $t \to \tilde{t} \tilde{N}$ with the top squarks'
decay being either semi-leptonic $\tilde{t} \to b \ell
\tilde\nu$ or flavor-changing $\tilde t \to c \tilde{N}, c \tilde{g}$.
Or maybe gluino pair production occurred, followed by $\tilde{g} \to t
\tilde{t}$.  These ideas can be tested using the rate, decay channels,
and kinematics of top quark events.\cite{symptom}

\subsection{Extra EW gauge bosons}\label{subsec:dewsb}

Many models of physics beyond the SM include extended electroweak
gague groups coupled differently to the third generation fermions
($h$) than to the light fermions ($\ell$).  Examples include
superstring theories with flavor U(1) groups and dynamical symmetry
breaking models with an extra SU(2) to produce $m_t$ or an extra U(1)
to explain $m_t - m_b$.

When the high-energy $SU(2)_h \times SU(2)_\ell \times U_1(Y)$ or
$SU(2)_W \times U(1)_h \times U(1)_\ell$ group breaks to
electromagnetism, the resulting gauge boson mass eigenstates are
\cite{ncetc} heavy states $W^H, Z^H$ that couple mainly to the third
generation, light states $W^L, Z^L$ resembling the standard $W$ and
$Z$, and a massless photon $A^\mu = \sin\theta [\sin\phi
\,W_{3\ell}^\mu + \cos\phi\,W_{3h}^\mu] +\cos\theta X^\mu $ coupling
to $Q = T_{3h} + T_{3\ell} + Y $.  Here, $\phi$ describes the mixing
between the original two weak or hypercharge groups and $\theta$ is
the usual weak angle.

There are several ways to test whether the high-energy weak
interactions have the form $SU(2)_h \times SU(2)_\ell$.  One
possibility is to search for the extra weak bosons.  Low-energy
exchange of $Z^H$ and $W^H$ bosons would cause apparent four-fermion
contact interactions; LEP limits on $eebb$ and $ee\tau\tau$ contact
terms imply\cite{zprimetautau} $M_{Z^H} \gae 400$ GeV.  Direct
production of $Z^H$ and $W^H$ at Fermilab is also feasible; a Run II
search for $Z^H \to \tau\tau \to e\mu X$ will be
sensitive\cite{zprimetautau} to $Z^H$ masses up to 750 GeV. Precision
electroweak data suggest that $M_{W^H,\, Z^H} \geq 2$ TeV if no other
physics cancels the bosons' effects.  Another possibility is to
measure the top quark's weak interactions in single top production.
Run II should measure the ratio of single top and single lepton
cross-sections $R_\sigma \equiv \sigma_{tb}/\sigma_{\ell\nu}$ to $\pm
8\%$ in the $W^*$ process.\cite{boos} making the experiments
sensitive\cite{testvtb} to $W^H$ bosons up to masses of about 1.5
TeV. 

There are likewise several ways to seek evidence for a $U(1)_h \times
U(1)_\ell$ structure.  CDF specifically \cite{leptophobe} excludes a
leptophobic topcolor $Z'$ decaying to $t\bar{t}$ if $M \leq$ 480 (780)
GeV assuming $\Gamma/M = 0.012\ (0.04)$.  Precision electroweak data
constrains\cite{tc2zp} flavor non-universal $Z'$ bosons to have masses
in excess of 1-2 TeV if no other physics cancels their
effects.  As mentioned earlier, FNAL Run II will be directly
sensitive\cite{zprimetautau} to $Z'$ bosons as heavy as 750 GeV via
$Z' \to \tau\tau \to e\mu X$.  A high-energy LC would be capable of
finding a 3-6 TeV $Z'$ decaying to taus.\cite{futc2}

\subsection{New Top Strong Interactions}\label{subsec:newstrong}

If the top quark feels a new strong interaction, a top condensate may
be involved in electroweak symmetry breaking. Consider, e.g., an
extended `topcolor' gauge structure\cite{tc2}: $SU(3)_h \times
SU(3)_\ell \to SU(3)_{QCD}$ where $t$ and $b$ transform under
$SU(3)_h$ and the light quarks, under $SU(3)_\ell$.  Below the
symmetry-breaking scale $M$, the spectrum includes massive topgluons
which mediate vectorial color-octet interactions among top quarks: $
-(4\pi\kappa/M^2) (\bar{t}\gamma_\mu \frac{\lambda^a}{2} t)^2$.  If
the coupling $\kappa$ lies above a critical value ($\kappa_c = 3\pi/8$
in the NJL\cite{NJL} approximation), a top condensate forms.  For a
second-order phase transition, $\langle \bar{t} t \rangle / M^3
\propto (\kappa - \kappa_c) / \kappa_c$, so the top quark mass
generated by this dynamics can lie well below the symmetry breaking
scale.

Models with strong top dynamics form three classes, whose 
distinctive spectra and phenomenology we now review:
topcolor,\cite{tc2,tc2phase} flavor-universal extended
color,\cite{futc2} and top seesaw.\cite{topseesaw}

\smallskip
\noindent{\tt Topcolor Models}
\smallskip

Topcolor\cite{tc2,tc2phase} models include extended color and
hypercharge sectors and a standard weak gauge group.  The
third-generation fermions transform under the more strongly-coupled
$SU(3)_h \times U(1)_h$ group, so that after the extended symmetry
breaks to the SM gauge group the heavy topgluons and $Z'$ couple
preferentially to the third generation.  The light fermions transform
under $SU(3)_\ell \times U(1)_\ell$.  CDF's search\cite{topglucdf} for
topgluons decaying to $b\bar{b}$ excludes 280 (340, 375) GeV $\leq M
\leq$ 670 (640, 560) GeV for $\Gamma/M$ = 0.3 (0.5, 0.7); the
topgluon's strong coupling to quarks makes it broad.  Run II and the LHC
should be sensitive to topgluons in $b\bar{b}$ or $t\bar{t}$ final
states. 

Because the topgluon and $Z'$ couple differently to quarks in
different generations, they contribute to flavor-changing neutral
currents.  Data on neutral B-meson mixing, e.g., implies
\cite{Burdman:2000in} $M_c >$ 3-5 TeV and $M_{Z'} >$ 7-10 TeV.

\begin{figure}[tb]
\begin{center}
\null\hspace{-1.5cm}
\rotatebox{90}{\scalebox{.33}{\includegraphics{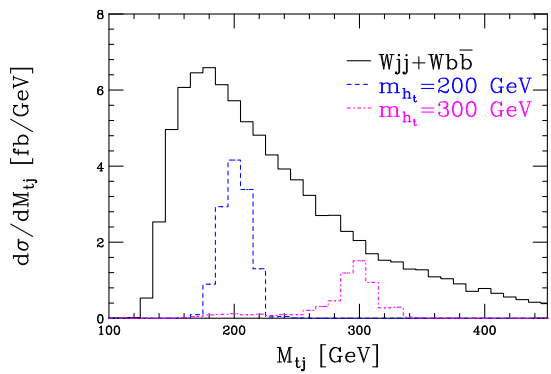}}}
\end{center}
\caption{ Simulated signal and background for charged top-pions in the
Tevatron's single top sample.\protect\cite{singtpion}}
\label{fig:toppi}
\end{figure}

The strong topcolor dynamics binds top and bottom quarks into a set of
top-pions\cite{tc2,tc2phase} $t\bar{t}, t\bar{b}, b\bar{t}$ and
$b\bar{b}$.  A singly-produced neutral top-higgs can be
detected\cite{fctpion} through its flavor-changing decays to $tc$ at
Run II.  Charged top-pions, on the other hand, would be
visible\cite{singtpion} in single top production, as in Figure
\ref{fig:toppi}, up to masses of 350 GeV at Run II and 1 TeV at LHC.

\smallskip
\noindent{\tt Flavor-Universal Coloron Models}
\smallskip

\begin{figure}[tb]
\begin{center}
\scalebox{.4}{\includegraphics{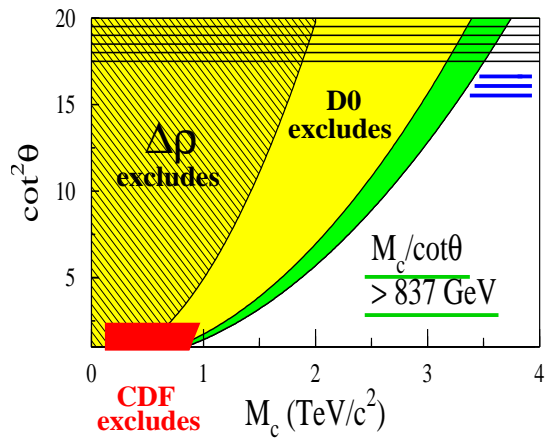}}
\end{center}
\caption{Limits\protect\cite{collim} on the mass and mixing angle of
flavor-universal colorons.}
\label{fig:coloron}
\end{figure}

The gauge structure of these models\cite{futc2} is identical to that
of topcolor\cite{tc2}.  The fermion hypercharges are as in
topcolor models; hence, the $Z'$ phenomenology is also the same. All
quarks transform under the more strongly-coupled $SU(3)_h$ group; none
transform under $SU(3)_\ell$.  As a result, the heavy coloron bosons
in the low-energy spectrum couple with equal strength to all quarks.
Experimental limits\cite{collim} on these color-octet states are shown
in Figure \ref{fig:coloron}. The most widely applicable bounds come
from the shape of the dijet angular distribution or invariant mass;
the current limit is $M_c / \cot\theta > 837$ GeV ($\theta$ is the
mixing angle between the $SU(3)$ groups).  This implies $M_c \gae 3.4$
TeV in dynamical models of mass generation where the coloron coupling
is strong.

These models are less constrained by flavor physics than topcolor models because both the colorons and composite pions are flavor-universal.  For example, B-meson mixing places only weak bounds on the $Z'$ mass \cite{Simmons:2001va}.

\smallskip
\noindent{\tt Top Seesaw Models}
\smallskip

\begin{table}[tb]
\caption{Third generation quark charge assignments in top seesaw
models.\protect\cite{topseesaw}}
\begin{center}
\begin{tabular}{|c|c|c|c|}\hline
 &  $ SU(3)_h$ & $ SU(3)_\ell$ &  $ SU(2)$ \\ \hline 
$ (t,\, b)_L$   &  3 &  1 &  2 \\
$ t_R,\ b_R$  &  1 &  3 &  1 \\
$ \chi_L$ &  1 &  3 &  1 \\
$ \chi_R$ &  3 &  1 &  1 \\
\hline
\end{tabular}
\end{center}
\label{tab:quarkchg}
\end{table}

Top seesaw models\cite{topseesaw} include an extended $SU(3)_h \times
SU(3)_\ell$ color group and a standard electroweak group.  In addition
to the ordinary quarks, new weak-singlet quarks
$\chi$  mix with top. The color and weak quantum numbers
of the third-generation quarks are shown in Table \ref{tab:quarkchg}.
When the $SU(3)_h$ coupling becomes strong, a dynamical top mass is created through a combination of $t_L \chi_R$
condensation and seesaw mixing:
\begin{equation}
\scalebox{.2}{\includegraphics{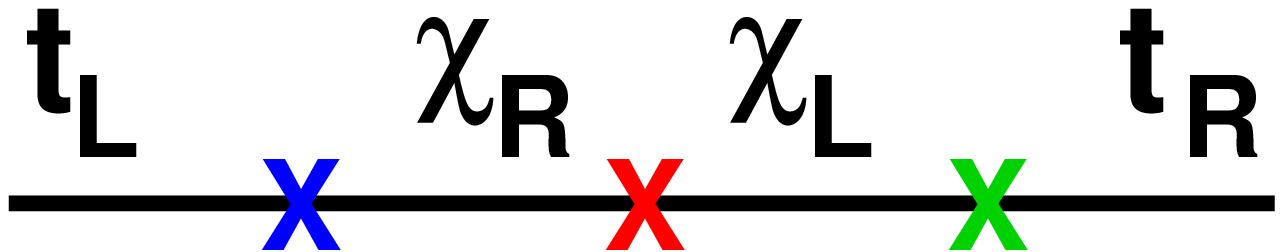}} \hspace{1cm}
\left( \begin{array}{cc} \bar{t}_L & \bar{\chi}_L \end{array} \right)
\left( \begin{array}{cc} 0 & {m_{t \chi}} \\ 
{\mu_{\chi t}} & {\mu_{\chi \chi}} \end{array} \right) 
\left( \begin{array}{c} t_R \\ \chi_R \end{array} \right)
\end{equation}
Composite scalars $\bar{t}_L \chi_R$ are also created by the strong
dynamics.

A combination of precision electroweak bounds and triviality
considerations limits the $\chi$ quarks to have masses $\gae$ 5
TeV. \cite{georgigrant,topseetriv} Direct searches for weak-singlet
quarks are limited to lower mass ranges, making them sensitive to
weak-singlet partners of the lighter quarks.  For example, a heavy
mostly-weak-singlet quark $q^H$ could contribute\cite{wsingferm} to
the FNAL top dilepton sample via $ p \bar{p} \to q^H \bar{q}^H \to q^L
W \bar{q}^L W \to q^L \bar{q}^L \ell \nu_\ell \ell' \nu_{\ell'}$ Run 1
data places the limit\cite{wsingferm} $M_{s^H, d^H} \gae 140$ GeV, but
cannot directly constrain $M_{b^H}$.  In models where all three
generations of quarks have weak-singlet partners, self-consistency
requires\cite{wsingferm} $M_{b^H}\gae 160$ GeV.

\section{Summary}\label{sec:summary-bsm}

The puzzles of electroweak symmetry breaking and fermion masses
require physics beyond the SM for their solution.  Thus, the top quark
may have visibly unusual attributes such as new gauge interactions or
decay channels, exotic fermion partners, a light supersymmetric
partner, or even strongly-bound top-quark states.  Our field needs new
surprising data, and hadron collider studies of top may soon provide
it!

\vspace{.5cm}
\begin{center}\textbf{Acknowledgments}\end{center}

\noindent This work was supported in part by the U.S. 
Department of Energy under grant DE-FG02-91ER40676

%

\end{document}